\documentclass[submission, Proceedings]{SciPost}
\usepackage{amsmath,amssymb,bm,graphicx,bbold,epsf,colordvi}

\hypersetup{
    colorlinks,
    linkcolor={red!50!black},
    citecolor={blue!50!black},
    urlcolor={blue!80!black}
}

\usepackage[bitstream-charter]{mathdesign}
\DeclareSymbolFont{usualmathcal}{OMS}{cmsy}{m}{n}
\DeclareSymbolFontAlphabet{\mathcal}{usualmathcal}

\usepackage{color}
\usepackage{mathrsfs}
\usepackage[export]{adjustbox}
\usepackage{accents}
\newcommand\thickbar[1]{\accentset{\rule{.4em}{.8pt}}{#1}}

\def\bea#1\eea{\begin{align}#1\end{align}}
\newcommand{\nn}{\nonumber}
\allowdisplaybreaks

\begin{document}

\begin{center}{\Large \textbf{
Resummation of the Sivers asymmetry in heavy flavor dijet production at the Electron-Ion Collider\\
}}\end{center}

\begin{center}
Z. B. Kang \textsuperscript{1,2,3},
J. Reiten \textsuperscript{1,2},
D. Y. Shao \textsuperscript{4,5}, and
J. Terry \textsuperscript{1,2$\star$}
\end{center}

\begin{center}
{\bf 1} Department of Physics and Astronomy, University of California, Los Angeles, USA
\\
{\bf 2} Mani L. Bhaumik Institute for Theoretical Physics, University of California, Los Angeles, USA
\\
{\bf 3} Center for Frontiers in Nuclear Science, Stony Brook University, Stony Brook, USA
\\
{\bf 4} Department of Physics and Center for Field Theory and Particle Physics, Fudan University, Shanghai, China
\\
{\bf 5} Key Laboratory of Nuclear Physics and Ion-beam Application (MOE), Fudan University, Shanghai, China
\\
* johndterry@physics.ucla.edu
\end{center}

\begin{center}
\today
\end{center}


\definecolor{palegray}{gray}{0.95}
\begin{center}
\colorbox{palegray}{
  \begin{tabular}{rr}
  \begin{minipage}{0.1\textwidth}
    \includegraphics[width=22mm]{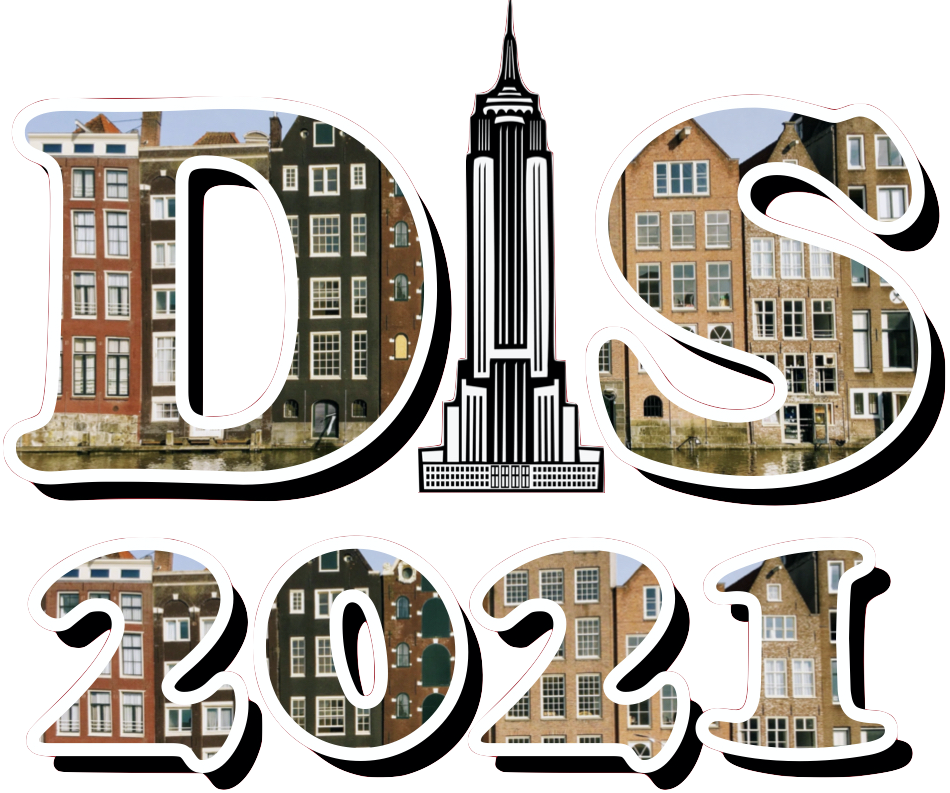}
  \end{minipage}
  &
  \begin{minipage}{0.75\textwidth}
    \begin{center}
    {\it Proceedings for the XXVIII International Workshop\\ on Deep-Inelastic Scattering and
Related Subjects,}\\
    {\it Stony Brook University, New York, USA, 12-16 April 2021} \\
    \doi{10.21468/SciPostPhysProc.?}\\
    \end{center}
  \end{minipage}
\end{tabular}
}
\end{center}

\section*{Abstract}
{\bf
We review our transverse momentum dependent factorization and resummation formalism for heavy flavor dijet production at the EIC. In this formalism, we have calculated the heavy flavor mass corrections in the collinear-soft and jet functions, and in the resummed expression for the cross section. By establishing this formalism, we then study the effects of the mass corrections by providing predictions at the EIC for the massive case and for the case where the mass is neglected. We find that the heavy flavor mass effects can give sizable corrections to the predicted asymmetry.
}

\section{Introduction}
Transverse-momentum-dependent parton distribution functions (TMD PDFs) encode the three-dimensional motion of confined partons in the nucleon. These distributions further also allow us to understand correlations between the parton's transverse momentum and the spin of parent nucleon~\cite{Accardi:2012qut}. The main focus of this study is the gluon Sivers function, which provides information on the correlation between a bound gluon's transverse momentum and the spin of the parent nucleon. A PYTHIA reweighting analysis was performed for the Sivers asymmetry in~\cite{Zheng:2018ssm} for open heavy flavor production, $D$ meson pair production, di-hadron production, and dijet production at the future Electron-Ion Collider (EIC). These authors found that the dijet process provided the cleanest probe of the parton level asymmetry. However, while dijet production at the EIC can be used to probe gluon TMD PDFs, this process also contains contamination from quark TMDs at leading order (LO). In this paper, we consider heavy flavor (HF) dijet production to further reduce background from the quark-initiated dijet process, see~\cite{Kang:2020xgk} for our full paper. In this paper, we establish the factorization and resummation formalism for the Sivers asymmetry in HF dijet production. Furthermore, we also study in detail the impact of the HF mass corrections.

\section{Factorization and resummation formula}\label{sec:fac}
We now review our factorization and resummation formalism for HF dijet production in polarized proton-electron scattering
\begin{align}\label{eq:pro}
    e(\ell) + p(P, \bm{S}_T) \to e(\ell') + J_{\mathcal{Q}}(p_J) + J_{\bar{\mathcal{Q}}}( p_{ \thickbar{J}}) + X\,.
\end{align}
For this process, $\bm{S}_T$ represents the transverse spin of the polarized proton while $\mathcal{Q}$ and $\bar{\mathcal{Q}}$ represent the HF quark and antiquark, which initiate the jet $J_{\mathcal{Q}}$ and $J_{\bar{\mathcal{Q}}}$ jets. For this analysis, we work in the Breit frame so that both the virtual photon and the beam proton move along the $z$-axis. In order to study TMDs, the measured total transverse momentum of the observed final state particles must be much smaller than the relevant hard scale for the interaction. For this process, two final state HF jets are detected. As a result, the transverse momentum scale is the dijet imbalance, which is defined as $\bm q_T = \bm p_{JT} +  {\bm p}_{\thickbar{J} T}$ where the $T$ subscript denotes the transverse component. The relevant hard scale for this process is the average transverse momentum of the jets. Therefore, the TMD region is valid for back-to-back jet production, ie $q_T \ll  {p}_{\thickbar{J} T} \sim p_{JT} \equiv p_T$. In addition, we also consider the HF mass, $m_{\mathcal{Q}}$, which introduces an additional scale. In this study, we focus on the kinematic region where $q_T \lesssim m_{\mathcal{Q}} \lesssim p_T R \ll p_T$ where $R$ denotes the jet radius. Using this power counting, the factorized expression for the unpolarized and polarized differential cross sections are given by
\begin{align}\label{eq:fac-unpl-rap}
    \frac{d\sigma^{UU}}{dQ^2 d y d^2 \bm p_{T} d y_J d^2 \bm q_T} = & H(Q,y,p_T,y_J,\mu)  \int \frac{d^2 b}{(2\pi)^2} e^{i\bm{b}\cdot \bm{q}_T} S(\bm b,\mu,\nu) \, f_{g/N}\left(x,b,\mu,\zeta/\nu^2\right) \,\notag \\
    & \hspace{-2.5cm} \times J_{\mathcal{Q}}(p_T R, m_{\mathcal{Q}}, \mu)\, S_{\mathcal{Q}}^{c}(\bm{b}, R, m_{\mathcal{Q}}, \mu)\, J_{\bar{\mathcal{Q}}}(p_T R, m_{\mathcal{Q}}, \mu)\, S_{\bar{\mathcal{Q}}}^{c}(\bm{b}, R, m_{\mathcal{Q}}, \mu) \,,
\end{align}
\begin{align}\label{eq:fac-poll-rap}
    \frac{d\sigma^{UT}(\bm S_T)}{dQ^2 d y d^2 \bm p_{T} d y_J d^2 \bm q_T} = & H(Q,y,p_T,y_J,\mu)  \int \frac{d^2 b}{(2\pi)^2} e^{i\bm{b}\cdot \bm{q}_T} S(\bm b,\mu,\nu) \nn \\
    & \hspace{-1.5cm} \times \frac{i}{2}\left(\epsilon_{\alpha\beta}\,S_T^\alpha\, b^\beta\right)\,f_{1T,g/N}^{\perp, f}(x,b,\mu,\zeta/\nu^2) \,\notag \\
    & \hspace{-1.5cm} \times J_{\mathcal{Q}}(p_T R, m_{\mathcal{Q}}, \mu)\, S_{\mathcal{Q}}^{c}(\bm{b}, R, m_{\mathcal{Q}}, \mu)\, J_{\bar{\mathcal{Q}}}(p_T R, m_{\mathcal{Q}}, \mu)\, S_{\bar{\mathcal{Q}}}^{c}(\bm{b}, R, m_{\mathcal{Q}}, \mu) \,.
\end{align}
In these expressions, $x$, $y$, and $Q^2$ are the usual DIS variables while $y_J$ is the rapidity of $J_{\mathcal{Q}}$. The superscripts of the $\sigma$ term denote the spin of the incoming lepton and proton, respectively, while $H$ is the hard function. We note that in general, the polarized process can have a different hard function than the unpolarized one ~\cite{Bacchetta:2005rm,Bomhof:2006dp,Kang:2020xez}. However, we find that the hard function should be equal for both processes for this process. The functions $J$ and $S^c$ are the jet and collinear-soft functions. We note that, due to the choice of the scale hierarchy, that both the jet and collinear-soft functions depend non-trivially on the mass of the HF quark. On the other hand, $S$ in Eqs.~\ref{eq:fac-unpl-rap} and~\ref{eq:fac-poll-rap} is the global soft function for the process while $f$ and $f_{1T}^\perp$ are the unpolarized gluon TMD PDF and the gluon Sivers function. We see that the $S$ as well as the TMD PDFs depend also on the rapidity scale $\nu$ due to the appearance of rapidity divergences. Finally, we also define $\mu$ as the renormalization scale while $\zeta$ is the Collin-Soper parameter.  

In order to generate the resummed expressions for the differential cross sections, each of the functions in the cross section must be evolved from their initial scale up to the hard scale by solving the RG equation $d F(\mu)/d \ln \mu = \Gamma_F(\alpha_s)\, F(\mu)$. Here $\Gamma_F$ represents the anomalous dimensions for the function $F$ which enter into Eqs.~\ref{eq:fac-unpl-rap} and ~\ref{eq:fac-poll-rap}. This procedure requires the explicit expression for the anomalous dimensions for each of the functions. In this paper, we provide the results of our calculations for the anomalous dimensions of $S$, $J$, and $S^c$ while we note that the anomalous dimensions of $H$ and the TMD PDFs are well-known, see for example \cite{Becher:2009th,Kang:2017glf}

\subsection{Anomalous Dimensions}\label{Sec:Hard}
By explicitly calculating the one loop expression for the global soft function, we find that the soft anomalous dimensions are given by
\begin{align}
    \label{eq:soft-ad}
    \Gamma^s(\alpha_s) & =  \gamma^{\rm{cusp}}(\alpha_s)\left[2 C_F\,\ln \frac{\mu^2}{\mu_b^2}  -C_A\,\ln \frac{\nu^2}{\mu^2}\right] + \frac{\alpha_s}{\pi}\left[2 C_F \ln\left( 4 c_{bJ}^2 \right) -\frac{1}{C_A}\ln\frac{\hat{s}}{p_T^2}+\,C_A \ln\frac{\hat{s}}{Q^2}\right]+ \mathcal{O}(\alpha_s^2)\,, 
    \\
    \label{eq:soft-rap-ad}
    \gamma_\nu^s(\alpha_s) &= -\frac{\alpha_s}{\pi}C_A \ln\frac{\mu^2}{\mu_b^2}+ \mathcal{O}(\alpha_s^2)\,.
\end{align}
In the top expression, $\hat{s}$ is the partonic COM energy, $c_{bJ} = \cos\left(\phi_b-\phi_J\right)$ where $\phi_{J(b)}$ are the azimuthal angle of the jet (impact parameter $\bm{b}$), and $\mu_b = 2 e^{-\gamma_E}/b$. In order to regulate the rapidity divergences of the soft function, we have used the $\eta$ regulator of \cite{Chiu:2012ir}. We note that the rapidity anomalous dimension of $S$ fulfills the RG consistency condition that  $\gamma_{\nu}^{f_g}+\gamma_{\nu}^s = 0$ where $\gamma_{\nu}^{f_g}$ is the one loop rapidity anomalous dimension of the gluon TMD PDF. Because of this relation, the product $f_{g/N}(x,b,\mu,\zeta/\nu^2)$ and $S(\bm b,\mu,\nu)$ is $\nu$-independent, and we can construct the properly-defined gluon TMD PDF
\begin{align}\label{e.prop_siv}
    f_{g/N}\left(x,b,\mu,\zeta/\nu^2\right) S(\bm b,\mu,\nu) = f_{g/N}^{\rm TMD}(x,b,\mu,\zeta) S(\bm b,\mu)\,.
\end{align} 

In our work, we have performed the explicit one loop calculation for the HF jet and collinear-soft functions and taken into consideration the mass dependence. We obtain the anomalous dimension for the jet and collinear-soft function to be
\begin{align}\label{eq:jf-ad}
    \Gamma^{j_{\mathcal{Q}}}(\alpha_s) &= - C_F \gamma^{\rm cusp}(\alpha_s) \ln \frac{p_T^2R^2\left(1+\beta\right)}{\mu^2} + \frac{\alpha_s C_F}{2\pi}\left( 3 - \frac{2\beta}{1+\beta} \right) + \mathcal{O}(\alpha_s^2)\,,
\\
\label{eq:cs-ad}
    \Gamma^{cs_{\mathcal{Q}}}(\alpha_s) &= C_F \gamma^{\rm cusp}(\alpha_s) \ln \frac{R^2 \mu_b^2}{\mu^2} + \frac{\alpha_s C_F}{\pi} \left[\ln\left(1+\beta\right) -2 \ln \left(-2i c_{bJ}\right) + \frac{\beta}{1+\beta} \right]+ \mathcal{O}(\alpha_s^2)\,,
\end{align}
where $\beta = m_{\mathcal{Q}}^2/p_T^2 R^2$. To arrive at Eqs.~\ref{eq:soft-ad}, \ref{eq:jf-ad}, and \ref{eq:cs-ad}, we have promoted $\alpha_s/\pi$ to $\gamma_{\rm cusp}$ in the log terms to obtain the anomalous dimensions at NLL accuracy.

\subsection{Resummation formula}\label{sec:xsec}
Having obtained the expressions for the anomalous dimensions, the resummed expressions for the differential cross section can be written as
\begin{align}\label{eq:res-unp}
    & \frac{d\sigma^{UU}}{dQ^2dyd^2\bm q_T dy_J d^2\bm p_T} = H(Q,y,p_T,y_J,\mu_h) \int_0^\infty \frac{b db}{2\pi}J_0(b\,q_T) f_{g/N}(x,\mu_{b*}) \notag \\
    &\hspace{0.25in} \times \exp \left[-\int_{\mu_{b *}}^{\mu_{h}} \frac{d \mu}{\mu} \Gamma^h\left(\alpha_{s}\right)-2\int_{\mu_{b *}}^{\mu_{j}} \frac{d \mu}{\mu} \Gamma^{j_{\mathcal{Q}}}\left(\alpha_{s}\right)-\int_{\mu_{b_{*}}}^{\mu_{c s}} \frac{d \mu}{\mu} \left(\bar\Gamma^{c s_{\mathcal{Q}}}\left(\alpha_{s}\right)+\bar \Gamma^{c s_{\bar{\mathcal{Q}}}}\left(\alpha_{s}\right)\right)\right]\notag \\
    &\hspace{0.25in} \times \exp\left[-S_{\rm NP}(b,Q_0,n\cdot p_g)\right]\,,
\\
\label{eq:res-siv}
    & \frac{d\sigma^{UT}(\bm S_T)}{dQ^2dyd^2\bm q_T dy_J d^2\bm p_T} = \sin(\phi_q-\phi_s )\, H(Q,y,p_T,y_J,\mu_h)\, \int_0^\infty \frac{b^2 db}{4\pi}J_1(b\,q_T) f_{1T, g/N}^{\perp, f}(x,\mu_{b*}) \notag \\
    &\hspace{0.25in}\times \exp \left[-\int_{\mu_{b *}}^{\mu_{h}} \frac{d \mu}{\mu} \Gamma^h\left(\alpha_{s}\right)-2\int_{\mu_{b *}}^{\mu_{j}} \frac{d \mu}{\mu} \Gamma^{j_{\mathcal{Q}}}\left(\alpha_{s}\right)-\int_{\mu_{b_{*}}}^{\mu_{c s}} \frac{d \mu}{\mu} \left(\bar\Gamma^{c s_{\mathcal{Q}}}\left(\alpha_{s}\right)+\bar\Gamma^{c s_{\bar{\mathcal{Q}}}}\left(\alpha_{s}\right)\right)\right]\notag \\
    &\hspace{0.25in}\times \exp\left[-S^\perp_{\rm NP}(b,Q_0,n\cdot p_g)\right]\,.
\end{align}
To arrive at these expressions, we have solved the RG evolution equations for each of the functions and used the RG consistency relation. In this expression, $\mu_h$, $\mu_j$, and $\mu_{cs}$ are the hard, jet, and collinear-soft scales, respectively. For the unpolarized gluon TMD PDF, we have performed LO collinear matching to the collinear gluon PDF $f_{g/N}(x,\mu)$ at the initial scales $\zeta_i = \mu_i^2 = \mu_{b*}^2$, where $\mu_{b*}$ is the natural scale obtained using the $b_*$-prescription \cite{Collins:1984kg}. The function $S_{\rm NP}$ in Eq.~\ref{eq:res-unp} parameterizes the non-perturbative dependence of the unpolarized TMD PDF. To obtain a parameterization for this function, we apply Casimir scaling to the parameterization given in \cite{Su:2014wpa}. For the gluon Sivers function, we adopt the non-perturbative parameterization utilized by \cite{DAlesio:2015fwo,Aschenauer:2015ndk}.

\section{Numerical results}\label{sec:num}
We select the energies of the electron and proton beam to be 20 GeV and 250 GeV, respectively. In Eqs.~\eqref{eq:res-unp} and \eqref{eq:res-siv} we take the scale choices $\mu_h = \sqrt{Q^2+p_T^2}$, $\mu_j=p_T R$, and $\mu_{cs}=\mu_{b_*}R$. Furthermore, we take the jet radius value $R=0.6$. For both charm and bottom dijet production, we take $|y_J|<4.5$. For charm jets, we also take $5\,{\rm GeV}<p_T<10\,{\rm GeV}$ and $m_c = 1.5$ GeV while for bottom dijets we take $10\,{\rm GeV}<p_T<15\,{\rm GeV}$ and $m_b = 5$ GeV. The spin asymmetry from the gluon Sivers function is defined as
\begin{align}
    A_{UT}^{\sin(\phi_q-\phi_s)} = 2 \frac{\int d\phi_s d\phi_q \sin(\phi_q-\phi_s) \, d\sigma^{UT}(\bm S_T)}{\int d\phi_s d\phi_q \, d\sigma^{UU} }\,.
\end{align}

\begin{figure}[htb!]
\begin{center}
\includegraphics[width=0.44\textwidth]{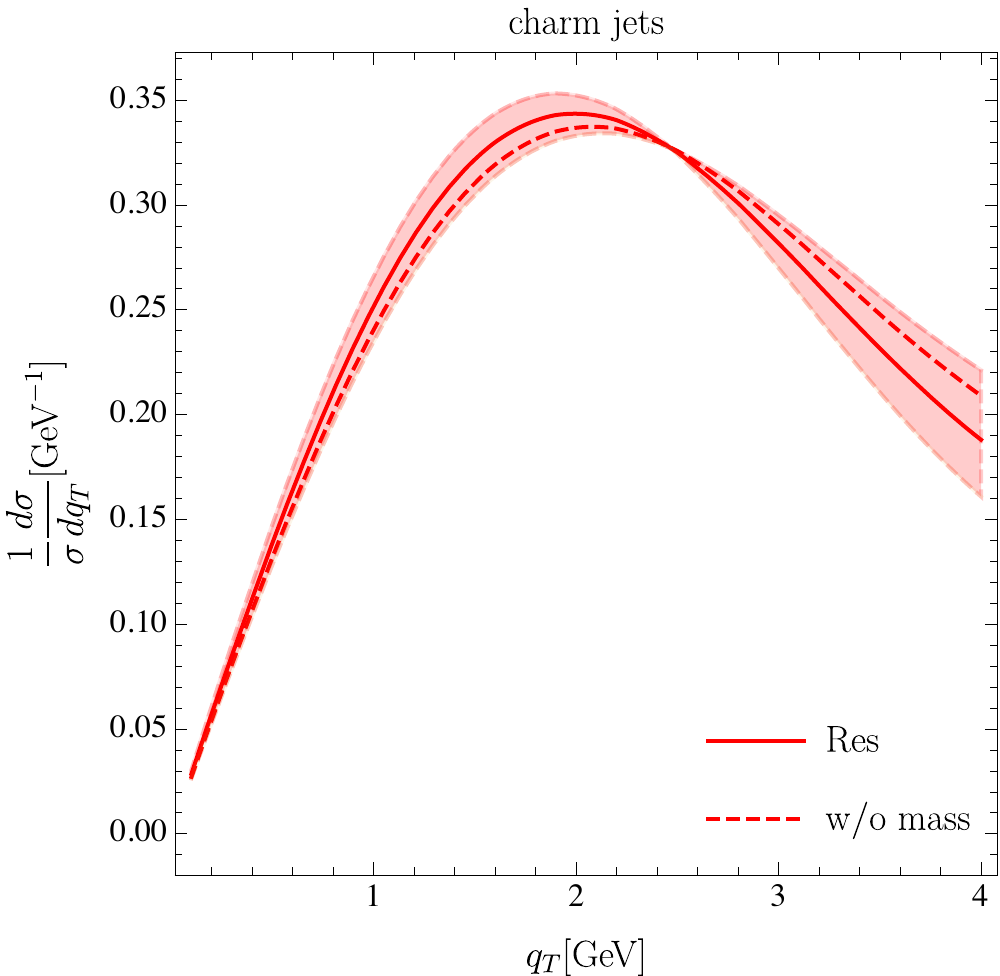}\hspace{0.5cm}
  \includegraphics[width=0.44\textwidth]{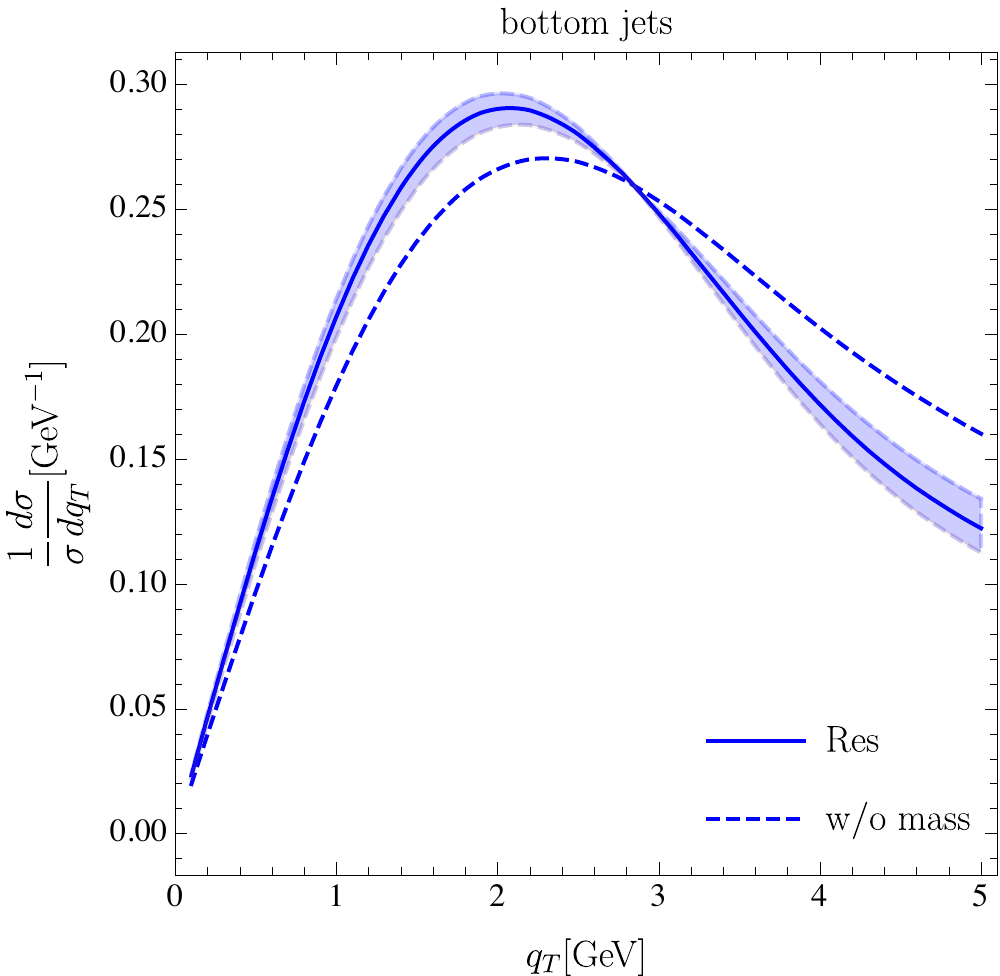}
\end{center}
  \caption{The $q_T$-distribution for the unpolarized cross section of charm (left) and bottom (right) dijet production at the EIC. The solid curves represent our prediction with mass corrections while we neglect mass corrections in the dashed curves. The bands represent our theoretical uncertainties obtained by varying the hard and jet scales.} 
\label{fig:pheno}
 \end{figure} 

\begin{figure}[hbt!]
\begin{center}
\includegraphics[width=0.44\textwidth]{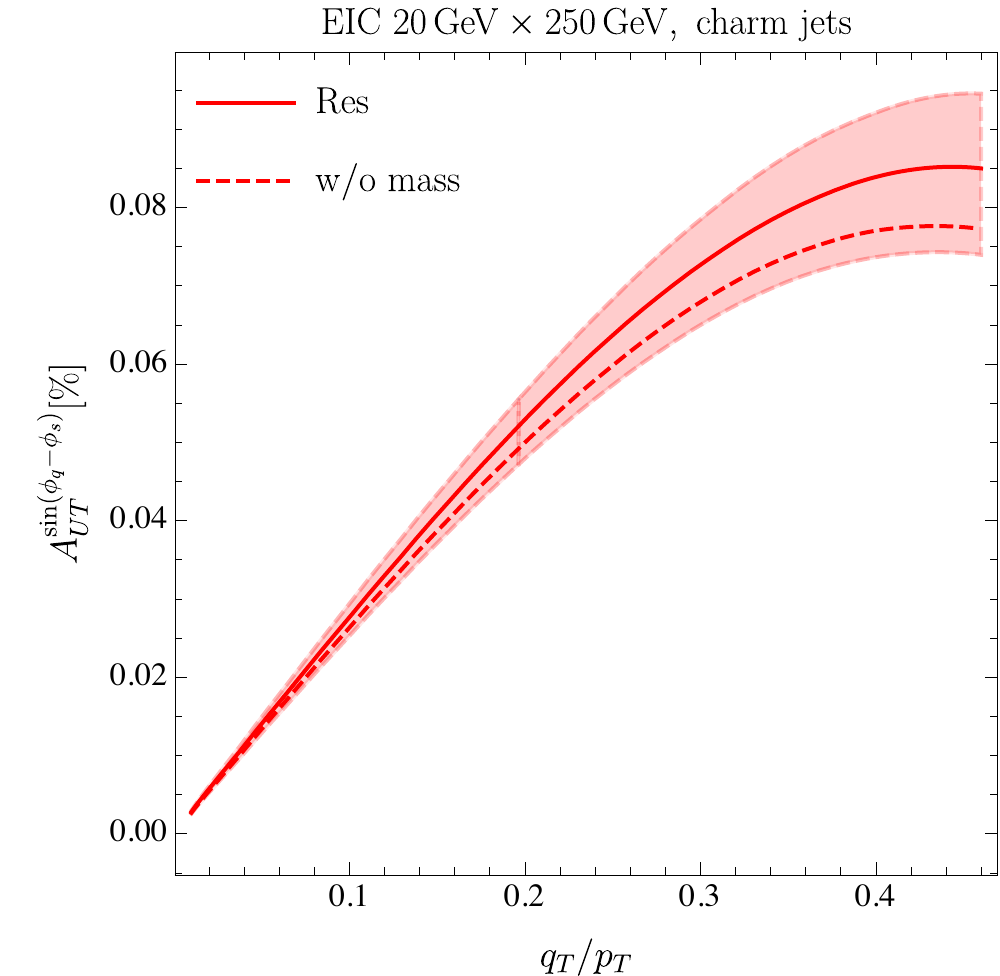}\hspace{0.5cm}
  \includegraphics[width=0.44\textwidth]{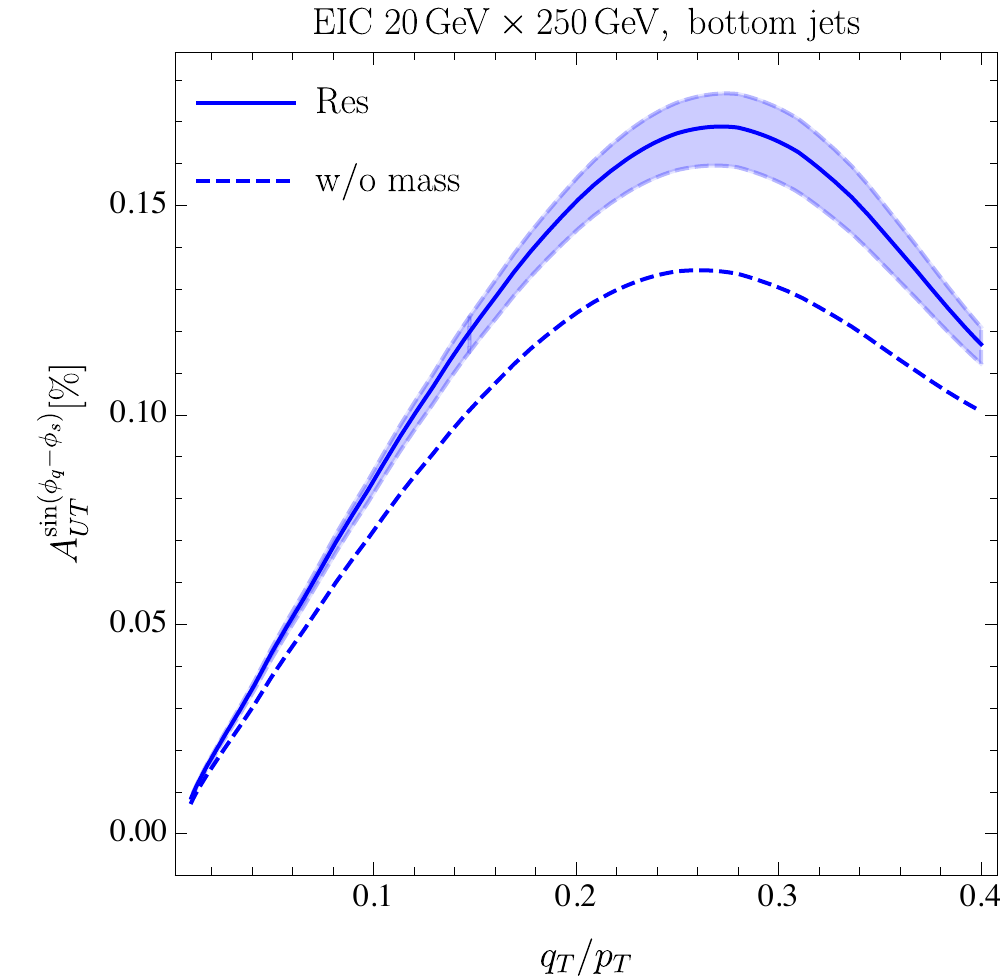}
\end{center}
  \caption{The Sivers asymmetry for charm dijet production (left) and bottom dijet production (right) as a function of $q_T/p_T$. The solid curves represent our prediction with mass corrections while we neglect mass corrections in the dashed curves. The bands represent our theoretical uncertainties obtained by varying the hard and jet scales.}
\label{fig:siv}
 \end{figure}

In Fig.~\ref{fig:pheno}, we provide the $q_T$ distribution for the unpolarized process. In Fig.~\ref{fig:siv}, the Sivers spin asymmetry $A_{UT}^{\sin(\phi_q-\phi_s)}$ is presented as a function of $q_T/p_T$ following~\cite{Arratia:2020nxw}, for both charm (left) and bottom (right) jets, respectively. In both plots, the solid curves represent the results obtained using the resummation formula with non-zero masses, while the dashed curves represent the resummation prediction using the evolution kernel without finite quark mass corrections. In Figs.~\ref{fig:pheno} and~\ref{fig:siv} we also provide the uncertainties originating from varying the hard and jet scales by a factor of two around their canonical values. We find that the size of the projected asymmetries are roughly 0.1$\%$. We also note that the size of our predicted asymmetry is consistent with the pseudo-data which was generated in~\cite{Zheng:2018ssm} for di-hadron production as well as dijet production. See the `SIDIS1' results of Figs.~11 and 14 of that paper for comparison. However, we note that the size of these projected asymmetry is heavily dependent on the non-perturbative parameterization used. 

\section{Conclusion}
In this work, we present our factorization and resummation formalism for the Sivers asymmetry in HF dijet production at the EIC. Using this formalism, we generate theoretical predictions with and without the HF mass corrections. We find in this analysis that the mass corrections for charmed dijet production are close in magnitude to the theoretical uncertainties from the scale variations. This indicates that a higher perturbative order is necessary in order resolve the mass effect. However, we find that the mass corrections are larger than the theoretical uncertainties for bottom dijet production. This indicates that the mass corrections could be important for future global analyses for bottom dijet production.

\paragraph{Funding information}
Z.K. and D.Y.S. are supported by the National Science Foundation under CAREER award PHY-1945471. J.R. is supported by the UC Office of the President through the UC Laboratory Fees Research Program under Grant No.~LGF-19-601097. J.T. is supported by NSF Graduate  Research Fellowship Program under Grant No. DGE-1650604. D.Y.S. is also supported by Center for Frontiers in Nuclear Science of Stony Brook University and Brookhaven National Laboratory. This work is supported within the framework of the TMD Topical Collaboration.

\bibliography{SciPost_Example_BiBTeX_File.bib}

\nolinenumbers

\end{document}